\documentclass[prb,preprint]{revtex4}

\newcommand{\beq}{\begin{equation}}
\newcommand{\eeq}{\end{equation}}
\newcommand{\beqar}{\begin{eqnarray}}
\newcommand{\eeqar}{\end{eqnarray}}

\ifx\undefined\fiverm
     \font\fiverm=cmr5
\fi

\input{prepictex}
\input{pictex}
\input{postpictex}

\begin{document}
\author{Tarun Biswas}
\title{Freely Falling Finite Frames Near a Black Hole}
\email{biswast@newpaltz.edu}
\affiliation{State University of New York at New Paltz, \\ New Paltz,  NY 12561, USA.}
\date{\today}
\begin{abstract}
It is well-known that the Riemann curvature tensor has no discontinuity at the black hole horizon.
It is also well-known that a freely falling observer takes finite time to reach the horizon
from an outside point. However, the usual assumption is that such an observer resides in a
frame of reference (spaceship) of infinitesimal size. This assumption is justified as long as the coordinates are continuous
enough to assume that the observer's frame is small compared to the variations of the metric
from a local flat metric. Such an assumption may be invalid when the coordinate system has not only a discontinuity
but a singularity like the one at the horizon. Hence, here, the characteristics of a finite frame (a spaceship)
near a black hole horizon is discussed. It is shown that clocks placed at the front and rear ends
have different time scales even in the limit when they reach the horizon at the same time. This
renders such a frame physically meaningless. It is also argued that
the forces that are expected to keep a realistic frame (like a spaceship) in one piece tend to
zero near the horizon. So, a physical spaceship is expected to fall apart near the horizon.
\end{abstract}
\maketitle
\section{Introduction}
A hundred years after Karl Schwarzschild's solution of Einstein's equations is a good time to
revisit his original solution\cite{schwarz,schwarzt,gron}. At present, the popularly accepted solution that is called the
Schwarzschild solution is actually a modification due to Droste and Hilbert\cite{gron,droste,hilbert}.
It is this modified version of the solution that opens up the possibility of the enigmatic black hole. Some have argued
against the existence of black holes for various reasons\cite{abrams,antoci1,antoci2}. In this paper, yet another
observational problem with black holes is discussed. Such collective evidence may require the return to the original
Schwarzschild solution.

It is commonly understood that an observer falling freely into a black hole does not notice
anything unusual while passing through the event horizon. This understanding is based on the
assumption that the observer's frame of reference (say a spaceship) is infinitesimal in size.
However, real frames can never be truly infinitesimal. So, one needs to investigate the validity
of the approximation of a finite frame by an infinitesimal one. The approximation should be valid
as long as the finite frame is small enough to keep variations of its metric from a local
flat metric negligible. This approximation can always be made if there are no discontinuities in
coordinates. In the presence of coordinate discontinuities, and in particular singularities,
more careful consideration is needed to see if the approximation is still valid. In particular,
the coordinate singularity at the event horizon deserves such careful consideration. Hence,
in the following, some features of a finite freely falling frame of reference will be considered.
To keep the discussion simple, the frame will be assumed to fall radially towards the black
hole.
\section{Difference in Time Measurements Between Front and Rear Ends of Falling Frame}
The Schwarzschild line element in standard spherical polar coordinates $(t,r,\theta,\phi)$ is given as
\beq
	d\tau^{2}=\left(1-\frac{r_{s}}{r}\right)dt^{2}-\left(1-\frac{r_{s}}{r}\right)^{-1}dr^{2}
	-r^{2}d\Omega^{2}, \label{eqmetric}
\eeq
where
\beq
	d\Omega^{2}=d\theta^{2}+\sin^{2}\theta d\phi^{2},
\eeq
the speed of light $c=1$, the Schwarzschild radius $r_{s}=2GM/c^{2}$, $G$ is the universal gravitational constant
and $M$ is the mass of the source. Using $\tau$ as the affine parameter, the two equations of motion of a radially freely
falling point particle are\cite{night,berg,misner},
\beqar
(1-r_{s}/r)^{-1}\frac{d^{2}r}{d\tau^{2}}+\frac{r_{s}}{2r^{2}}\left(\frac{dt}{d\tau}\right)^{2} & & \nonumber \\
-(1-r_{s}/r)^{-2}\frac{r_{s}}{2r^{2}}\left(\frac{dr}{d\tau}\right)^{2} & = & 0, \\
\frac{d}{d\tau}\left((1-r_{s}/r)\frac{dt}{d\tau}\right) & = & 0.
\eeqar
On integration, these give,
\beqar
\left(\frac{dr}{d\tau}\right)^{2} & = & k^{2}-(1-r_{s}/r), \label{eqradial} \\
(1-r_{s}/r)\frac{dt}{d\tau} & = & k, \label{eqtime}
\eeqar
where $k$ is a constant that depends on initial conditions.
If the falling particle starts at $r=r_{0}$ at zero velocity, then equation~\ref{eqradial}
gives,
\beq
k=(1-r_{s}/r_{0})^{1/2}.
\eeq
A more compact way of writing equations~\ref{eqradial} and~\ref{eqtime} is as follows.
\beqar
dr & = & -\sqrt{k^{2}-A}\frac{A}{k}dt, \label{eqradialc} \\
d\tau & = & \frac{A}{k}dt, \label{eqtimec}
\eeqar
where,
\beq
A=1-r_{s}/r.
\eeq
The negative sign for the square root is chosen as $dr/dt$ is negative for a falling particle.

Now, consider a finite sized frame of reference (spaceship) falling freely along a radial
direction towards the event horizon at $r=r_{s}$. Let the front end of the spaceship be
given by $r=r_{1}$ and the rear end be given by $r=r_{2}$ at any time $t$. Also, let the
frame start from rest at $r_{1}=r_{01}$ and $r_{2}=r_{02}$. A measure of the initial length of the spaceship is,
\beq
L=r_{02}-r_{01}.\label{eqsslength}
\eeq
For a frame of finite length, $L$ must be finite.
As we are interested primarily in the behavior of the spaceship near the black hole horizon, we can assume
the non-gravitational forces keeping the spaceship together to be small compared to the tidal forces. Hence, each of the two
ends can be thought of as falling freely. So, the equations~\ref{eqradialc} and~\ref{eqtimec} can be
written for each of the two ends as follows (using the subindices `1' and `2' for the front and rear ends respectively).
\beqar
dr_{1} & = & -\sqrt{k_{1}^{2}-A_{1}}\frac{A_{1}}{k_{1}}dt, \label{eqradial1} \\
d\tau_{1} & = & \frac{A_{1}}{k_{1}}dt, \label{eqtime1} \\
dr_{2} & = & -\sqrt{k_{2}^{2}-A_{2}}\frac{A_{2}}{k_{2}}dt, \label{eqradial2} \\
d\tau_{2} & = & \frac{A_{2}}{k_{2}}dt \label{eqtime2},
\eeqar
where,
\beqar
k_{1}=(1-r_{s}/r_{01})^{1/2}, & \; & A_{1}=1-r_{s}/r_{1}, \\
k_{2}=(1-r_{s}/r_{02})^{1/2}, & \; & A_{2}=1-r_{s}/r_{2}.
\eeqar
If a clock placed at the front end measures the proper time interval $d\tau_{1}$, it is the
invariant magnitude of the space-time separation between two events given by $(t, r_{1})$ and $(t+dt, r_{1}+dr_{1})$
where $dr_{1}$ is the change in the position of the falling front end in time $dt$.
Let us now compute the time interval of the same two events as measured by a clock placed at
the rear end. First, we transform $dr_{1}$ and $dt$ to the stationary, locally flat coordinate system
at the instantaneous position of the rear end. This requires the metric scale factors as shown in
equation~\ref{eqmetric}. So, the space and time intervals in this frame are,
\beqar
dr' & = & A_{2}^{-1/2}dr_{1}, \label{eqrprime} \\
dt' & = & A_{2}^{1/2}dt. \label{eqtprime}
\eeqar
Next, this needs to be transformed to the moving frame of the rear end using a Lorentz boost due to the
velocity $v_{2}$ of the rear end with respect to the local stationary frame. This gives,
\beqar
dR & = & \frac{dr'- v_{2}dt'}{\sqrt{1-v_{2}^{2}}}, \label{eqrearr1}\\
dT & = & \frac{dt'- v_{2}dr'}{\sqrt{1-v_{2}^{2}}}, \label{eqreart1}
\eeqar
where $dR$ is the radial component and $dT$ the time component. The velocity of a falling particle with respect
to the stationary, locally flat frame can be found (using the metric scaling of equation~\ref{eqmetric}) to be,
\beq
v=\frac{A^{-1/2}dr}{A^{1/2}dt} = \frac{1}{A}\frac{dr}{dt}.
\eeq
Using equation~\ref{eqradialc}, this gives,
\beq
v = -\sqrt{k^{2}-A}/k. \label{eqrelv}
\eeq
Specifically for the rear end this gives,
\beq
v_{2} = -\sqrt{k_{2}^{2}-A_{2}}/k_{2}. \label{eqrelv2}
\eeq
Using this in equation~\ref{eqreart1} and making substitutions from equations~\ref{eqrprime}, \ref{eqtprime} and~\ref{eqradial1}
gives,
\beq
dT = \left(k_{2}-\frac{A_{1}\sqrt{(k_{1}^{2}-A_{1})(k_{2}^{2}-A_{2})}}{k_{1}A_{2}}\right)dt.
\eeq
Next, converting $dt$ to $d\tau_{1}$ using equations~\ref{eqtime1} gives,
\beq
dT = \left(\frac{k_{1}k_{2}}{A_{1}}-\frac{\sqrt{(k_{1}^{2}-A_{1})(k_{2}^{2}-A_{2})}}{A_{2}}\right)d\tau_{1}.
\eeq
So the time interval in consideration is measured at the rear end to be different from the front
end measurement by a scale factor $S$.
\beq
dT = Sd\tau_{1},
\eeq
where,
\beq
S = \left(\frac{k_{1}k_{2}}{A_{1}}-\frac{\sqrt{(k_{1}^{2}-A_{1})(k_{2}^{2}-A_{2})}}{A_{2}}\right).
\label{eqscale}
\eeq
To test the integrity of the finite frame at the event horizon, we need to find $S$ in the
limit of the spaceship
reaching the horizon. This is the limit of $t\to\infty$, and hence, $r_{1}\to r_{s}$ and
$r_{2}\to r_{s}$. So, in this limit $A_{1}\to 0$ and $A_{2}\to 0$.
As a result, we see that the limiting value of the scale factor,
\beq
S_{0}=\lim_{t\to\infty}S,
\eeq
has an indeterminate form. So, the limit has to be computed carefully. This has been done in
appendix~\ref{app} and the result is as follows.
\beq
S_{0} = \frac{k_{1}^{2}+k_{2}^{2}}{2k_{1}k_{2}}.
\eeq
Note that $S_{0}=1$, if $k_{1}=k_{2}$. This is expected, as $k_{1}=k_{2}$ means both front and
rear ends start at the same place making them the same frame. However, if $k_{1}\neq k_{2}$,
$S_{0}>1$. This means that, at the horizon, the front and rear end time measurements for the
same space-time interval will be different by a finite factor. Note that the two ends coincide on
reaching the horizon. Also, using equation~\ref{eqrelv}, it can be seen that both ends travel at the
speed of light at the horizon. Hence, their time measurements being different by a finite factor
is meaningless. So, finite frames of reference are meaningless at the event horizon.

One may also consider the extreme limit of the front end starting at the horizon ($r_{01}=r_{s}$).
In this case, as long as the rear end starts at a finite distance from the front, the scale factor
$S_{0}$ becomes infinity\footnote{This limit has to be interpreted only as a limit. If $r_{01}=r_{s}$
exactly, the procedure of appendix~\ref{app} fails.}.\addtocounter{footnote}{-1}
This is definitely not acceptable for any frame of reference!

\section{Relative Velocity of Front and Rear Ends}
Now, let us compute the relative velocity of the front end with respect to the rear end. This would be
$ dR/dT $. $ dR $ is computed using equations~\ref{eqrprime}, \ref{eqtprime}, \ref{eqrearr1}, \ref{eqrelv2}
and also~\ref{eqradial1} and~\ref{eqtime1}. The result is,
\beq
dR=S_{r}d\tau_{1},
\eeq
where,
\beq
S_{r}=\frac{k_{1}\sqrt{k_{2}^{2}-A_{2}}}{A_{1}} - \frac{k_{2}\sqrt{k_{1}^{2}-A_{1}}}{A_{2}}. \label{eqsr}
\eeq
Then the relative velocity is,
\beq
V =\frac{dR}{dT}=\frac{S_{r}}{S}.
\eeq
In the limit of the frame reaching the horizon, $ S_{r} $ is found to be (see appendix~\ref{app}),
\beq
S_{r0}=\lim_{t\to\infty}S_{r}=\frac{k_{2}^{2}-k_{1}^{2}}{2k_{1}k_{2}}. \label{eqsr0}
\eeq
Hence, the limiting case of the relative velocity of the front end with respect to the rear end is,
\beq
V_{0}=\lim_{t\to\infty}V=\frac{k_{2}^{2}-k_{1}^{2}}{k_{1}^{2}+k_{2}^{2}}.
\eeq
For $ k_{1}\neq k_{2} $, it can be seen that $ V_{0}>0 $ as $ k_{2}>k_{1} $. A positive relative
velocity using the radial coordinate means the front end is moving outwards and hence, towards the
rear end. So, tidal forces near the horizon will tend to compress the finite frame. This contradicts
the common understanding derived from the weak gravity limit (Newtonian gravity).

In the extreme limit of the front end starting at the horizon ($ r_{01}=r_{s} $), $ V_{0} $ becomes
unity which is the speed of light\footnotemark.

\section{Forces Near the Event Horizon}
A physical spaceship, in order to stay in one piece, relies on non-gravitational forces.
A non-gravitational force can communicate between the ends of the
spaceship at most at the speed of light. The above results show that there can be a finite
time delay in information transmitted from the front to the rear even when the two ends are only
infinitesimally apart. This means the effective speed of light tends to zero and communication
of forces between the ends fails. Hence, at the horizon,
the spaceship is expected to come apart.

\section{Conclusion}
Here it has been shown that a finite frame of reference, like a spaceship, becomes meaningless
when it reaches the event horizon of a black hole. Its front and rear ends reach the horizon
together and at the same speed. But their time measurements of a space-time interval are different
by a finite factor. Also, it is argued that forces that keep a spaceship together fail to transmit
between the ends. As a result, the spaceship comes apart at the horizon.
This, and other unphysical aspects of the black hole event horizon\cite{abrams,antoci1,antoci2,bis1},
make the existence of such an event horizon suspect. A possible resolution is presented in an earlier publication\cite{bis2}.

\appendix
\section{Limiting Values of Scale Factors}
\label{app}
Equation~\ref{eqscale} shows that $S$ takes an indeterminate form as $t \to \infty$. As $A_{1}$
and $A_{2}$ tend to zero in this limit, one can expand the square roots in powers of $A_{1}/k_{1}^{2}$
and $A_{2}/k_{2}^{2}$ and keep up to first order terms. This gives,
\beq
S \simeq k_{1}k_{2}\left(\frac{1}{A_{1}}-\frac{1}{A_{2}}\right)+\frac{k_{2}A_{1}}{2k_{1}A_{2}}
+\frac{k_{1}}{2k_{2}}. \label{eqsapprox}
\eeq
This still contains indeterminate forms. So, we need to find the relationship of $ A_{1} $ and
$ A_{2} $ near the horizon. This can, of course, be found by integrating equations~\ref{eqradial1}
and~\ref{eqradial2}. But a simpler approach is to realize that in the limit of $ A_{1}\to 0 $
and $ A_{2}\to 0 $, they can be related as follows.
\beq
A_{2}=\alpha A_{1}^{p},
\eeq
where $ \alpha $ and $ p $ are positive constants. Now, noting that $ A_{2}\geq A_{1} $,
it can be seen that $ S\to\infty $ as
$ A_{1}\to 0 $ unless $ \alpha=1 $ and $ p=1 $. This is true even for $ k_{1}=k_{2} $. But,
we must have $ S=1 $ for $ k_{1}=k_{2} $ as that is the case of the front and rear ends being
the same throughout their trajectories. Hence, we conclude that $ \alpha=p=1 $. This gives,
\beq
A_{2}=A_{1},
\eeq
in the limit $ A_{1}\to 0 $. Hence, in the same limit, we get (using equation~\ref{eqsapprox}),
\beq
S_{0}=\lim_{t\to\infty}S=\frac{k_{1}^{2}+k_{2}^{2}}{2k_{1}k_{2}}.
\eeq

Similarly, $ S_{r0} $ of equation~\ref{eqsr0} can be computed from equation~\ref{eqsr}. The result is,
\beq
S_{r0}=\frac{k_{2}^{2}-k_{1}^{2}}{2k_{1}k_{2}}.
\eeq

\end{document}